\documentclass[12pt]{article}

\usepackage[utf8]{inputenc}
\usepackage[T1]{fontenc}

\usepackage{setspace} 
\usepackage[vmargin = 1.5in, hmargin = 1.5in]{geometry}

\usepackage[usenames,dvipsnames,svgnames]{xcolor}
\usepackage{amsmath, amssymb, amsfonts, graphicx, tikz,pdflscape, mathtools, amsthm, upgreek, bm,pgfplots,csvsimple,multirow, multicol, booktabs}
\usepackage{verbatim}
\usepackage{natbib}
\usepackage{accents}
\newcommand{\ubar}[1]{\underaccent{\bar}{#1}}
\usepackage{needspace}
\def\citeapos#1{\citeauthor{#1}'s (\citeyear{#1})}
\usepackage{comment}
\usepackage[inline]{enumitem}
\usepackage{tocvsec2}
\usepackage{threeparttable}
\usetikzlibrary{calc, cd}
\usepackage[skip = 10pt]{caption}
\allowdisplaybreaks[1]
\allowdisplaybreaks[1]

\definecolor{Blue}{RGB}{86,180,233}
\definecolor{Orange}{RGB}{230,159,0}
\definecolor{Green}{RGB}{0,158,115}
\definecolor{GmailBlue}{RGB}{42, 93, 176} 
\usepackage[
pagebackref,
colorlinks=true,
citecolor= GmailBlue,
linkcolor=GmailBlue,
urlcolor = GmailBlue
]{hyperref}

\newcommand{\bibtexorder}[1]{}

\usepackage{pgfplots}
\pgfplotsset{compat=newest}
\usepgfplotslibrary{groupplots}
\usepackage{tikz}
\usetikzlibrary{matrix,calc,shapes,arrows.meta,positioning}
\pgfplotsset{width = \textwidth/2}
\tikzstyle{hollow}=[circle,draw,inner sep=1.5]
\tikzstyle{solid}=[circle,draw,inner sep=1.5,fill=black]
\pgfplotsset{compat = newest}

\usepackage[capitalize,noabbrev]{cleveref}

\newtheorem{thm}{Theorem}
\newtheorem*{theorem*}{Theorem}
\newtheorem*{cor*}{Corollary}
\newtheorem{cor}{Corollary}

\newtheorem{lem}{Lemma}

\crefname{prop}{Proposition}{Propositions}
\crefname{thm}{Theorem}{Theorems}
\crefname{lem}{Lemma}{Lemmas}

\crefname{as}{Assumption}{Assumptions}

\theoremstyle{definition}

\newtheorem*{rem*}{Remark}

\crefname{lem2}{Lemma}{Lemmas}

\def\a{\alpha}
\def\b{\beta}

\def\e{\varepsilon}

\def\l{\lambda}

\label{key}

\def\D{\Delta}

\def\S{\Sigma}

\def\R{\mathbf{R}}

\def\FF{\mathcal{F}}

\DeclareMathOperator{\spn}{span} 

\newcommand{\ang}[1]{\langle #1 \rangle} 

\newcommand{\paren}[1]{( #1 )} 

\newcommand{\Paren}[1]{\left( #1 \right)}

\newcommand{\Brac}[1]{\left[ #1 \right]}

\newcommand{\Set}[1]{\left\{ #1 \right\}}

\tikzcdset{every label/.append style = {font = \small}}

\title{Simple Proofs of the  Variational and Multiple Priors Representations\thanks{This paper subsumes my previous note titled ``A Simpler Proof of Gilboa and Schmeidler’s (1989) Maxmin Expected Utility.'' I thank Roberto Corrao for bringing to my attention \citeapos{Marinacci1998} note.}}
\author{Ian Ball\thanks{MIT Department of Economics, \texttt{ianball@mit.edu}.}}
\date{14 April 2024}

\begin{document}
	
\maketitle

This note gives a simpler proof of the variational representation in \cite{MMR}.  The main tool is the (algebraic) Hahn--Banach theorem. No other results from functional analysis are needed. In fact, the proof does not require a topology on the space of acts. 

Next, I apply a construction from the main proof to give a short proof of \citeapos{CMMRstructure} characterization of the structure of variational preferences. Finally, I follow the argument in the variational proof to give a simple proof of the multiple priors representation in \cite{GS}. In this case, the key step is similar to a result in \cite{Marinacci1998}.

\section{Variational and multiple  priors preferences}

\subsection{Set-up}

Consider the Anscombe--Aumann setting of \cite{MMR}. There is a  set $S$ of states of the world, endowed with an algebra $\S$ of events. The set $X$ of consequences is a convex subset of a vector space. The set $\FF$ of acts consists of all simple $\S$-measurable functions $f \colon S \to X$. Convex combinations of acts are evaluated pointwise. Each consequence $x$ in $X$ is identified with the constant act that equals $x$ in every state. In this way, we view $X$ as a subset of $\FF$. Preferences over acts are given by a binary relation $\succsim$ on $\FF$. Let  $\succ$ and $\sim$ denote the asymmetric and symmetric parts of $\succsim$. Write $f \precsim g$ if and only if $g \succsim f$.

\subsection{Axioms}

The following axioms appear in \cite{MMR} and \cite{GS}. 

\begin{enumerate}
	\item[A1.\hphantom{$^\prime$}] \emph{Nontrivial weak order}: $\succsim$ is complete, transitive, and nontrivial.\footnote{A binary relation $\succsim$ on $\FF$ is trivial if $f \succsim g$ for all $f,g \in \FF$.}
	\item[A2.\hphantom{$^\prime$}] \emph{Monotonicity}: For all $f,g \in \FF$, if $f(s) \succsim g(s)$ for all $s \in S$, then $f \succsim g$. 
	\item[A3.\hphantom{$^\prime$}] \emph{Mixture continuity}: For all $f,g,h \in \FF$, the sets $\{ \a  \in [0,1]: \a f + (1 - \a) g \succsim h\}$ and $\{ \a  \in [0,1]:   h \succsim \a f + (1 - \a) g\}$ are closed.
	\item [A3$^\prime$.] \emph{Archimedean}: For all $f,g,h \in \FF$, if $f \succ g$ and $g \succ h$, then there exist $\a, \b \in (0,1)$ such that $\a f + (1 - \a) h \succ g$ and $g \succ \b f + (1 - \b) h$. 
	\item[A4.\hphantom{$^\prime$}] \emph{Weak certainty independence}: For all $f,g \in \FF$, $x,y \in X$, and $\a \in (0,1)$, if $\a f + (1 - \a) x \succsim a g + (1 - \a ) x$, then $\a f + (1 - \a) y \succsim a g + (1 - \a ) y$. 
	\item[A4$^\prime$.] \emph{Certainty independence}: For all $f,g \in \FF$, $x \in X$, and $\a \in (0,1)$, if $f \succ g$, then $\a f + (1 - \a) x \succ \a g + (1 - \a ) x$.\footnote{The original version in \cite{GS} requires ``if and only if.'' }
	\item[A5.\hphantom{$^\prime$}]	\emph{Uncertainty aversion}: For all $f,g \in \FF$ and $\a \in (0,1)$,  if $f \sim g$ then $\a f +(1 - \a) g \succsim f$.
\end{enumerate}

\subsection{Representation theorems}

A function $V \colon \FF \to \R$ \emph{represents} a binary relation $\succsim$ on $\FF$ if we have
\[
	f \succsim g \iff V(f) \geq V(g), \qquad f,g \in \FF.
\]
Let $B_0$ denote the space of  simple $\S$-measurable functions $\varphi \colon S \to \R$. Let $\D$ denote the set of probabilities, i.e., finitely additive probability measures on $(S, \S)$. For any function $\varphi$ in $B_0$ and any probability $p$ in $\D$, denote by $\ang{\varphi, p}$ the integral of $\varphi$ against $p$. The space $\D$ is endowed with the weak topology induced by the maps $\ang{ \varphi, \cdot }$ for each $\varphi$ in $B_0$. A function on a convex set is \emph{affine} if it is both convex and concave. A function is \emph{grounded} if its infimum is $0$.

\begin{thm}[Variational representation; \citet{MMR}] \label{res:MMR}  \label{res:representation_varitional} For a binary relation $\succsim$ on $\FF$, the following are equivalent:
\begin{enumerate}
	\item $\succsim$ satisfies $\mathrm{A1}, \mathrm{A2}, \mathrm{A3}, \mathrm{A4}, \mathrm{A5}$;
	\item There exists a nonconstant affine  function $u \colon X \to \R$ and a grounded, lower semicontinuous, convex function $c \colon \D\to [0,\infty]$ such that $\succsim$ is represented by the function $V \colon \FF \to \R$ defined by
	\[
	V(f) = \min_{p \in \D} \Brac{ \ang{u \circ f, p} + c(p) }.
	\]
\end{enumerate}
\end{thm}

\begin{thm}[Multiple priors representation; \cite{GS}] \label{res:GS} \label{res:representation_multiple_priors} For a binary relation $\succsim$ on $\FF$, the following are equivalent:
	\begin{enumerate}
		\item $\succsim$ satisfies $\mathrm{A1}, \mathrm{A2}, \mathrm{A3}^\prime, \mathrm{A4}^\prime, \mathrm{A5}$;
		\item There exists a nonconstant affine function $u \colon X \to \R$ and a nonempty,  closed, convex subset $C$ of $\D$ such that $\succsim$ is represented by the function $V \colon \FF \to \R$ defined by
		\[
		V(f) = \min_{p \in C} \, \ang{u \circ f, p}.
		\]
	\end{enumerate}
\end{thm}

\subsection{Notation for the proofs}

For any subset $K$ of $\R$,  let $B_0(K)$ denote the set of simple $\S$-measurable functions $\varphi \colon S \to K$. A functional $J \colon B_0(K) \to \R$ is 
\begin{itemize}
	\item \emph{normalized} if $J (k  1_S) = k$ for all $k$ in $K$;
	\item \emph{mixture continuous} if for all $\varphi,\psi \in B_0(K)$, the map $t \mapsto J(t \varphi + (1 - t) \psi)$ is continuous on $[0,1]$;
	\item \emph{vertically invariant} if $J ( \a \varphi + (1 - \a)  k 1_S) =  J (\a \varphi) + (1 - \a) k$, for all $\varphi \in B_0(K)$, $k \in K$, and $\a \in (0,1)$;
	\item 	\emph{certainty independent} if $J ( \a \varphi + (1 - \a)  k 1_S) =  \a J (\varphi) + (1 - \a) k$, for all $\varphi \in B_0(K)$, $k \in K$, and $\a \in (0,1)$. 
\end{itemize}
Below, we often denote the constant function $k 1_S$ simply as $k$. 

\section{Discussion of the proofs}

For \cref{res:MMR} and \cref{res:GS}, I prove the main implication from the axioms to the representation. Both proofs have the following structure. 
\begin{enumerate}
	\item Use  the restricted form of the independence axiom to show that the restriction of $\succsim$ to $X$ has an affine representation $u$ on $X$. 
	\item Extend $u$ to a representation $V$ on $\FF$, which can be shown to have the form $V(f) = I (u \circ f)$ for some functional $I \colon B_0 (u(X)) \to \R$.
	\item Use the axioms to establish properties of the functional $I$.  In particular, the functional $I$ is normalized, monotone, and concave. 
	\item Express the functional $I$ as the minimum of a class of dominating affine functionals. In the case of variational preferences, this class consists of \emph{translations} of  monotone,  normalized  linear functionals. In the case of multiple priors preferences, this class consists of \emph{untranslated} monotone,  normalized linear functionals. Almost by definition, monotone,  normalized linear functionals  correspond to integration against finitely additive probability measures. 
\end{enumerate}

The proofs lay bare the role of finitely additive probability measures in the multiple priors and variational representations. These probabilities are associated with dominating affine functionals on the space of utility-valued acts. The original proofs, by contrast, use more tools from functional analysis. Those proofs pass to the supnorm closure of $B_0$, which is a Banach space. The norm dual space has a classical representation from functional analysis. The proof of the variational representation, which is split between \cite{MMR} and \cite{CMMRniveloids}, uses quite abstract mathematical tools, namely, the theory of niveloids on Archimedean Riesz spaces. 

I hope that the more elementary proofs in this note make these fundamental representation theorems more accessible.

\section{Proof of variational representation}

We prove the forward implication. 


\begin{lem} The restriction of~$\mathord{\succsim}$ to $X$ has an affine representation $u \colon X \to \R$ satisfying $u(X) \supseteq [-1,1]$.
\end{lem}

\begin{proof} We show that the restriction of $\mathord{\succsim}$ to $X$ satisfies the independence axiom in the mixture space theorem \citep{HersteinMilnor1953}. Fix $x,y,z \in X$ with $x \sim y$. We claim that $\frac{1}{2} x + \frac{1}{2} z \sim \frac{1}{2}y + \frac{1}{2} z$. If not, we may assume without loss that $\frac{1}{2} x + \frac{1}{2} z \succ \frac{1}{2} y + \frac{1}{2} z$. Applying weak certainty independence twice with $x$ in place of $z$ and then twice with $y$ in place of $z$ gives $x \succ \frac{1}{2}x + \frac{1}{2} y \succ y$, which is a contradiction. So $\mathbin{\succsim}$ is represented on $X$ by an affine utility function $u \colon X \to \R$. This function $u$ cannot be constant; otherwise, by monotonicity, $\succsim$ would be trivial on $\FF$.  The range $u(X)$ is convex, so after scaling and translating $u$, we may assume that $u(X) \supseteq [-1,1]$.
\end{proof}

We claim that every act has a certainty equivalent. For any act $f$, there exist $\bar{x}, \ubar{x} \in X$ such that $\bar{x} \succsim f(s) \succsim \ubar{x}$ for every $s$. Using mixture continuity and the connectedness of $[0,1]$, it can be shown that $f \sim \a \ubar{x} + (1 - \a) \bar{x}$ for some $\a$ in $[0,1]$.

Define $V \colon \FF \to \R$ by $V(f) = u(x_f)$, where $x_f$ is a certainty equivalent of $f$. 
The value of $u(x_f)$ does not depend on the choice of certainty equivalent. Note that $ \{ u \circ f:  f \in \FF\} = B_0 (u(X))$. Define $I \colon B_0( u(X)) \to \R$ by $I ( u \circ f) = u(x_f)$. This functional is well-defined by monotonicity: if $u \circ f = u \circ g$, then $f \sim g$. To prove properties of $I$, it is convenient to define a preference relation $\succsim'$ on $B_0 ( u(X))$ by setting $u \circ f \succsim' u \circ g$ iff $f \succsim g$. Thus, $I$ represents $\succsim'$. 

Viewing $K$ as a subset of $B_0(K)$, observe that $\succsim'$ reduces to $\geq$ on $K$. Since $u$ is affine, $\succsim'$ inherits axioms A1--A5 from $\succsim$ (with $K$ in place of $X$). 

\begin{lem} The functional $I \colon B_0 (u(X)) \to \R$ is  normalized, monotone, mixture continuous, vertically invariant, and concave. 
\end{lem}

\begin{proof} 	For each $x \in X$, we have $ I (u (x) 1_S) = I(u \circ x) = u(x)$, so $I$ is normalized.  Monotonicity and mixture continuity follow from the corresponding properties of $\succsim'$.

	
	We next prove vertical invariance. Fix $\varphi \in B_0 (u(X))$, $k \in u(X)$, and $\a \in (0,1)$. Since $I$ is normalized,
	\begin{equation} \label{eq:indifference}
		\a \varphi + (1 - \a) 0  = \a \varphi \sim' I ( \a \varphi) =  \a \paren{ \a^{-1} I(\a \varphi)} + (1 - \a) 0.
	\end{equation}
	By monotonicity, we have $\a^{-1} I (\a \varphi)\in [ \min \varphi, \max \varphi] \subseteq u(X)$, so we can apply weak certainty independence to conclude that
	\[
	\a \varphi + (1 - \a) k \sim' \a ( \a^{-1} I( \a \varphi)) + (1 - \a) k = I(\a \varphi) + (1 - \a) k.
	\]
	Thus, $I ( \a \varphi + (1- \a) k) = I( \a \varphi) + (1- \a) k$. 
		
	For concavity,  we first prove \emph{weak concavity}: for all $\e  > 0 $ and all $\varphi, \psi \in B_0 ( (1 - \e) u(X))$ satisfying $|I(\varphi) - I(\psi) | \leq \e$, we have 
	\begin{equation} \label{eq:weak_concave}
		I ( \a \varphi + (1 - \a) \psi) \geq \a I (\varphi) + (1 - \a) I(\psi), \qquad \a \in (0,1).
	\end{equation}
	To prove this, let $r = I(\psi) - I(\varphi)$ and let $\varphi' = \varphi + r$. Since $|r| \leq \e$, we know that $r$ is in $\e u(X)$. By vertical invariance, $I (\varphi') = I (\varphi) + r = I(\psi)$. For all $\a \in (0,1)$, uncertainty aversion implies that
	\begin{equation} \label{eq:plus_r}
		I(\psi) \leq I ( \a \varphi' + (1 - \a) \psi) = I ( \a \varphi + (1 - \a) \psi)+ \a r,
	\end{equation}
	where the equality follows from vertical invariance since  $\a \varphi + (1 - \a) \psi$ is in $B_0 ( (1 - \e) u(X))$ and $\a r$ is in $\e u(X)$. Subtract $\a r$ from both sides of \eqref{eq:plus_r} to get \eqref{eq:weak_concave}.
	
	Now we show that $I$ is concave. Fix $\varphi, \psi \in B_0 (K)$ and $\a \in (0,1)$. For each $t \in (0,1)$, the function $I$ is locally concave on the segment $[t \varphi, t \psi]$ (by weak concavity and mixture continuity) and hence concave, by a result from convex analysis.\footnote{Consider a function $f \colon [0,1] \to \R$. For $0 \leq x < y \leq 1$, say that $f$ violates concavity on the secant of  $[x,y]$ if there exists $\a$ in $(0,1)$ such that $f(  \a x + (1 - \a) y) < \a f(x) + (1 - \a) f(y)$. It can be shown that for any nondegenerate interval $[x,y]$, if  $f$ violates concavity on the secant of $[x,y]$, then $f$ violates concavity on the secant of  at least one of the following intervals: $[x, (x + y)/2]$, $[(x + y)/2, y]$, and $[ (3x + y)/4, (x + 3 y)/4]$. By successively applying this argument,  we can show that if $f$ is not concave, then $f$ violates concavity on the secants of a nested sequence of intervals whose intersection is a single point. This contradicts local concavity.} Thus,
	\[
	I \bigl( \a (t \varphi) + (1 - \a) (t \psi) \bigr) \geq \a I (t \varphi ) + (1  -\a)  I ( t \psi).
	\]
	Pass to the limit as $t \uparrow 1$. By mixture continuity, $I ( \a \varphi + (1 - \a) \psi) \geq \a I ( \varphi) + (1 -\a) I (\psi)$. 
\end{proof}

\begin{lem} \label{res:variational_3} Let $K$ be a convex subset of $\R$ satisfying $K \supseteq [-1,1]$. If a functional $I \colon B_0 (K) \to \R$ is normalized, monotone, mixture continuous, vertically invariant, and concave, then there exists a grounded, lower semicontinuous, convex function $c \colon \D \to [0,\infty]$ such that
\begin{equation} \label{eq:variational_eq}
	I( \varphi) = \min_{p \in  \D} \Set{ \ang{\varphi,p} + c(p) },\qquad \varphi \in B_0(K). 
\end{equation}
\end{lem}

	\begin{proof}  Fix  $\varphi \in B_0( \operatorname{int} K)$. Since $I$ is concave, the Hahn--Banach separation theorem ensures that there exists a linear functional $\ell$ on $B_0$ such that $\ell ( \psi -  \varphi) \geq I(\psi) - I(\varphi)$ for all $ \psi \in B_0 (K)$. For any $\psi' \in B_0$, we know that for $\e$ sufficiently small, $\varphi + \e \psi'$ is in $B_0(K)$, and hence
	\[
	\e \ell (\psi') = \ell (\e \psi') \geq I ( \varphi  +\e \psi') - I(\varphi).
	\]
	Thus, for nonnegative  $\psi'$ in $B_0$, we have $\ell(\psi') \geq 0$ by the monotonicity of $I$. 
To see that $\ell$ is normalized, take $\psi' = \pm 1 \in K$, and apply vertical invariance, noting that $\varphi$ is in $B_0( (1 -\e)K)$ for $\e$ sufficiently small. We conclude that $\ell = \ang{ \cdot, p_\varphi}$ for some probability $p_\varphi$ in $\D$.
	
	For each $p \in \D$, let $c(p) = \sup_{\varphi \in B_0(K)}  \Paren{ I(\varphi) - \ang{\varphi,p}}$.  The function $c$ is lower semicontinuous and convex (being the supremum over continuous linear functionals). If $c$ satisfies \eqref{eq:variational_eq} with $\varphi = 0$, then $c$ must be grounded. So it suffices to prove \eqref{eq:variational_eq}. By construction, $\min_{p \in \D} \Paren{ \ang{\varphi, p} + c(p) } \geq I (\varphi)$ for all $\varphi \in B_0(K)$. At each $\varphi$ in $B_0 (\operatorname{int} K)$, we have equality by the construction of $p_\varphi$ above. To see that equality holds at any $\varphi \in B_0(K)$, note that for each $t$ in $(0,1)$, we have
	\begin{equation*}
		\begin{aligned}
			\ang{ \varphi, p_{t\varphi}} + c( t \varphi) - I(\varphi) 
			&= \ang{ (1 -t) \varphi, p_{t\varphi}} +  I(t \varphi) - I(\varphi) \\
			&\leq (1 -t) \max \varphi + \Paren{ I(t \varphi) - I(\varphi)}.
		\end{aligned}
	\end{equation*}
	As $t \uparrow 1$, the right side tends to $0$ by mixture continuity.
\end{proof}

\section{Structure of variational preferences}

We now generalize the construction in the proof of \cref{res:variational_3} in order to characterize the class of cost functions that satisfy \eqref{eq:variational_eq}. For any functional  $J \colon B_0 \to \R$, define $J^\dagger \colon \D \to (-\infty , \infty]$ by $J^\dagger (p) = \sup_{\varphi \in B_0}  \Brac{ J (\varphi) - \ang{ \varphi, p}}$. For any functional $a \colon \D \to \R$, define $a^\ast \colon B_0 \to \R$ by $a^\ast (\varphi) = \min_{p \in \D} \Brac{ \ang{ \varphi, p} + a(p)}$.   These operations are monotone. 

Let $I \colon B_0 (K) \to \R$ be a functional of the form in \cref{res:variational_3}. Extend $I$ to a functional $\ubar{I} \colon B_0 \to [-\infty, \infty)$ by setting $\ubar{I}( \varphi) = - \infty$ for $\varphi \notin B_0 (K)$. In terms of the operators defined above, the proof of \cref{res:variational_3} takes $c = \ubar{I}^\dagger$, and then shows that $c^\ast$ agrees with $I$ on $B_0(K)$.

We could have instead proven \cref{res:variational_3} with a different cost function construction. Define $\bar{I} \colon B_0 \to \R$ by 
\begin{equation} \label{eq:barI}
	\bar{I} ( \psi)  = \inf_{p}  \Paren{ \ang {\psi, p} + \ubar{I}^\dagger (p)},
\end{equation}
where the infimum is over every probability $p$ in $\Delta$ that is a supergradient of $I$ at some point in $B_0 (\operatorname{int} K)$. With $c = \bar{I}^\dagger$, the proof can be completed as before since for each $\varphi$ in $B_0( \operatorname{int} K)$, the supergradient $p_\varphi$ of $I$ constructed above is also a supergradient of $\bar{I}$.

A grounded, lower semicontinuous, convex function $c \colon \D \to [0,\infty]$ \emph{represents} $I$ if \eqref{eq:variational_eq} holds, i.e.,  if $c^\ast$ agrees with $I$ on $B_0(K)$.  We can now give a very short proof of a result from \cite{CMMRstructure} on the structure of the class of cost functions that represent $I$.  

\begin{thm}[Cost structure]  \label{res:structure} Let $I \colon B_0(K) \to \R$ be of the form in \cref{res:variational_3}. Let  $\ubar{c} = \ubar{I}^\dagger$ and $\bar{c} = \bar{I}^\dagger$. 
	\begin{enumerate} 
		\item  \label{it:inequality} If a  concave functional $J \colon B_0 \to \bar{\R}$ extends $I$, then $\ubar{I} \leq J \leq \bar{I}$. 
		\item A convex, grounded, lower semicontinuous function $c \colon \D \to [0,\infty]$ represents $I$ if and only if $\ubar{c} \leq c \leq \bar{c}$.
\end{enumerate}
\end{thm}

\begin{proof} 
1.  Clearly, $\ubar{I} \leq J$.  We show that $J \leq \bar{I}$. If not,  then $J(\psi) > \bar{I}(\psi)$ for some $\psi$ in $B_0$. By the definition of $\bar{I}$, there exists a probability $p$ in $\D$ that is a supergradient of  $I$ at some point $\varphi$ in $B_0 ( \operatorname{int} K)$ such that $J(\psi) > \ang{\psi, p} + \ubar{I}^\dagger (p)$.  It can be shown that $J$ is strictly larger than $I$ on the interval $(\varphi, \psi] \cap B_0 (K)$, which is a contradiction.

2. If $\ubar{c} \leq c \leq \bar{c}$, then $\ubar{c}^\ast \leq c^\ast \leq \bar{c}^\ast$. As argued above, $\ubar{c}^\ast$ and $\bar{c}^\ast$ agree with $I$ on $B_0(K)$,  so $c^\ast$ does too. Conversely, if $c$ represents $I$, then by  part \ref{it:inequality},  we have  $\ubar{I} \leq c^\ast  \leq \bar{I}$. Thus, $\ubar{c}= \ubar{I}^\dagger \leq c^{\ast \dagger} \leq \bar{I}^\dagger = \bar{c}$. By the Fenchel-Moreau theorem,\footnote{Observe that $-J^\dagger$ is the restriction to $\D$ of the Fenchel conjugate of $J$. For $a \colon \D \to [0, \infty]$, let $\bar{a}$ be the extension of $a$ to $\operatorname{ ba} (S, \S) = \spn \D$ defined by $\bar{a}( p) = \infty$ for $p \not\in \D$. Then $-a^\ast$ is the Fenchel conjugate of $\bar{a}$. Therefore, $a^{\ast \dagger}$ is the restriction to $\D$ of the Fenchel biconjugate of $\bar{a}$.} $c^{\ast \dagger} = c$. 
\end{proof}

\begin{cor}[Uniqueness] Let $I \colon B_0 (K) \to \R$ be of the form in \cref{res:variational_3}. If $K$ is unbounded, then exactly one grounded, lower semicontinuous, convex function $c \colon \D \to [0,\infty]$ represents $I$.
\end{cor}

\begin{proof} By \cref{res:structure}, it suffices to show that $\bar{I}^\dagger \leq \ubar{I}^\dagger$. From \eqref{eq:barI}, it is clear that $\bar{I} (\varphi + k) = \bar{I}(\varphi) + k$, for all $\varphi$ in $B_0$ and $k$ in $\R$. If  $K$ is unbounded, then for each $\varphi \in B_0$, there exists $k$ in $\R$ such that $\varphi + k \in B_0(K)$. Hence,
\begin{equation}
	\bar{I} (\varphi) - \ang{\varphi, p} 
	= \bar{I} ( \varphi + k) - \ang{\varphi + k, p} 
	\leq \ubar{I}^{\dagger} (p),
\end{equation}
where the inequality holds because $\bar{I}$ agrees with $I$ on $B_0(K)$. Take the supremum of the left side over $\varphi$ in $B_0$ to conclude that $\bar{I}^\dagger (p) \leq \ubar{I}^\dagger (p)$. 
\end{proof}

\section{Proof of multiple priors representation}

We prove the forward implication. 

\begin{lem} The restriction of~$\mathop{\succsim}$ to $X$ has an affine representation $u \colon X \to \R$ satisfying $u(X) \supseteq [-1,1]$.
\end{lem}

\begin{proof} By certainty independence, the restriction of $\succsim$ to $X$ satisfies the von Neumann--Morgenstern axioms, and hence is represented by an affine utility function $u \colon X \to \R$.\footnote{See, e.g.,  \citet[Theorem 8.4, p.~112--113]{Fishburn}.} This function $u$ cannot be constant; otherwise, by monotonicity, $\succsim$ would be trivial on $\FF$.  The range $u(X)$ is convex, so after scaling and translating $u$, we may assume that $u(X) \supseteq [-1,1]$.
\end{proof}

We claim that every act has a certainty equivalent. For any act $f$, there exist $\bar{x}, \ubar{x} \in X$ such that $\bar{x} \succsim f(s) \succsim \ubar{x}$ for every $s$. Using mixture continuity and the connectedness of $[0,1]$, it can be shown that $f \sim \a \ubar{x} + (1 - \a) \bar{x}$ for some $\a$ in $[0,1]$.

Define $V \colon \FF \to \R$ by $V(f) = u(x_f)$, where $x_f$ is a certainty equivalent of $f$. 
The value of $u(x_f)$ does not depend on the choice of certainty equivalent. Note that $ \{ u \circ f:  f \in \FF\} = B_0 (u(X))$. Define $I \colon B_0( u(X)) \to \R$ by $I ( u \circ f) = u(x_f)$. This functional is well-defined by monotonicity: if $u \circ f = u \circ g$, then $f \sim g$. To prove properties of $I$, it is convenient to define a preference relation $\succsim'$ on $B_0 ( u(X))$ by setting $u \circ f \succsim' u \circ g$ iff $f \succsim g$. Thus, $I$ represents $\succsim'$. 

Viewing $K$ as a subset of $B_0(K)$, observe that $\succsim'$ reduces to $\geq$ on $K$. Since $u$ is affine, $\succsim'$ inherits axioms $\mathrm{A1}, \mathrm{A2}, \mathrm{A3}^\prime, \mathrm{A4}^\prime, \mathrm{A5}$ from $\succsim$ (with $K$ in place of $X$).

\begin{lem}  \label{lem:superlinear} The functional $I \colon B_0 ( u(X)) \to \R$ is normalized, monotone, and superlinear. 
\end{lem}

\begin{proof} For each $x \in X$, we have $ I (u (x) 1_S) = I(u \circ x) = u(x)$, so $I$ is normalized.  Monotonicity follows from the monotonicity of $\succsim'$. 
	
To prove superlinearity, we first prove that $I$ is certainty independent. Fix $\varphi  \in B_0 (K)$, $k \in u(X)$, and $\a \in (0,1)$. Since $I$ is normalized, $\varphi \sim' I(\varphi)$. Since $\succsim'$ is certainty independent, $\a \varphi  + (1 -\a) k \sim' \a I(\varphi) + (1 -\a) k$, hence $I ( \a \varphi + (1-\a) k) = \a I(\varphi) +(1-\a) k$. Positive homogeneity is immediate from certainty independence.\footnote{If $\varphi = \l \psi$ for $\l > 1$ then $\psi = \l^{-1} \varphi$ and $\l^{-1} < 1$.}

Now we prove that $I$ is superadditive. By positive homogeneity, it suffices to show that $I( \frac{1}{2} \varphi + \frac{1}{2}\psi) \geq \frac{1}{2} I(\varphi) + \frac{1}{2} I(\psi)$ for any $\varphi,\psi \in B_0(\frac{1}{2}u(X))$ for which $r \coloneqq I(\psi) - I(\varphi)$ is in $\frac{1}{2} u(X)$. To prove this property, let $\varphi' = \varphi + r$. Since $I$ is certainty independent, $I(\varphi') = I(\varphi) + r = I(\psi)$, so by uncertainty aversion, $I(\psi) \leq I ( \frac{1}{2} \varphi' + \frac{1}{2} \psi) = I  ( \frac{1}{2} \varphi + \frac{1}{2} \psi)  + \frac{1}{2}r$, where the equality follows from certainty independence. Subtract $\frac{1}{2}r$ from both sides.
\end{proof}

\begin{lem} \label{res:multiple_prior_affine} Let $K$ be a convex subset of $\R$ satisfying $K \supseteq [-1,1]$. If a  functional $I \colon B_0( K) \to \R$ is normalized, monotone, and superlinear, then there exists a  nonempty,  closed, convex subset $C$ of $\D$ such that $I(\varphi) = \min_{p \in C}\, \ang{\varphi, p}$ for all $\varphi$ in $B_0(K)$.
\end{lem}

\begin{proof} Using positive homogeneity, we extend $I$ to a normalized, monotone,  superlinear functional on $B_0$.  For each $\varphi$ in $B_0$ the restriction of $I$ to $\spn \varphi$ is a linear functional. By the Hahn--Banach extension theorem, this linear functional can be extended to a linear functional  $\ell$ on $B_0$ that dominates $I$ (and agrees with $I$ on $\spn \varphi $).  Since $I$ is monotone, $\ell$ is nonnegative: $\psi \geq 0$ implies $\ell(\psi) \geq I(\psi) \geq I(0) = 0$. Moreover, $\ell$ is normalized since $\pm \ell(1) = \ell ( \pm 1) \geq I ( \pm 1) = \pm 1$.  We conclude that $\ell = \ang{ \cdot, p_\varphi}$ for some probability $p_{\varphi}$ in $\D$. Let $C$ be the set of all probabilities $p$ in $\D$  satisfying $\ang{\psi, p} \geq I(\psi)$ for all $\psi$ in $B_0(K)$. Note that $C$ is nonempty, closed, and convex. The supergradient construction above shows that $I( \varphi) =  \ang{\varphi, p_\varphi} = \min_{p \in C} \, \ang{\varphi, p}$ for all $\varphi$ in $B_0(K)$. 
\end{proof}

\cref{res:multiple_prior_affine} is similar to the result in \cite{Marinacci1998}.

\newpage
\bibliographystyle{ecta}
\bibliography{GS_lit.bib}

\end{document}